# An Integrated Roadside Sensing and Communication Framework for Vulnerable Road User Safety at Signalized Intersections

Parvez Anowar[1*]

[1]PhD Student, Department of Civil, Environmental and Construction Engineering, University of Central Florida, 12800 Pegasus Drive, Orlando, FL 32816, United States

## Abstract

Vulnerable road users (VRUs) account for approximately half of urban traffic deaths globally, with intersections concentrating a disproportionate share of these casualties. Recent reviews of sensing technology for VRU protection have cataloged dozens of single-sensor and dual-sensor deployments, yet none of the surveyed systems couples multi-modal sensing with edge-side near-miss analytics and bidirectional vehicle-to-everything (V2X) and pedestrian-to-everything (P2X) messaging in a single intersection cabinet. This paper presents an integrated framework for VRU protection at signalized intersections, combining LiDAR, radar, RGB camera, and thermal camera at the perception layer, edge-based prediction and surrogate-safety analytics at the computation layer, V2X and P2X messaging at the communication layer, and adaptive signal control at the actuation layer. The framework is grounded in an empirical case study using R-LiViT, the first publicly released roadside LiDAR-Visual-Thermal dataset, which provides 200 multi-modal sequences and 2,400 annotated RGB-T frames at three German intersections. Analysis of 53,319 detection annotations reveals that VRUs comprise approximately 49% of all road-user observations, that day-to-night density drops by 38% for pedestrians and 45% for vehicles while the night distribution shows a higher close-proximity share, that per-frame close-proximity event counts vary approximately 10-fold across the eight unique locations at three intersections, and that 83% of pedestrian bounding boxes are small in image space, indicating that VRUs are typically far from any single sensor. These findings support multi-modal sensing, edge-side analytics, and adaptive context-sensitive deployment rather than uniform single-sensor solutions.

## Key findings

- VRUs account for approximately 49% of road-user detections at three intersections.
- Day-to-night pedestrian density drops 38%, but VRU-vehicle pairs sit closer at night.
- Close-proximity events per frame vary nearly 10-fold across eight locations.
- 83% of pedestrian boxes are small, indicating distant single-sensor detection.
- Multi-modal sensing, edge analytics, and V2X/P2X are not yet integrated in practice.

* Corresponding author: Parvez Anowar (pa545735@ucf.edu)

## 1. Introduction

Pedestrians, cyclists, motorcyclists, and powered two- and three-wheeler riders, collectively termed vulnerable road users (VRUs), bear a disproportionate burden of road traffic mortality. The World Health Organization reports that pedestrians, cyclists, and powered two- and three-wheeler riders together account for approximately 56% of global road traffic deaths, with 21% attributed to pedestrians, 30% to powered two- and three-wheeler riders, and 5% to cyclists, summarised by the WHO as roughly half of all road traffic fatalities (World Health Organization, 2023). Intersections host roughly half of traffic injuries and more than 2.8 million crashes annually in the United States (Mir et al., 2025).

In-vehicle countermeasures such as pedestrian-detection emergency braking and forward collision warning mitigate risk only after a hazard enters the host vehicle's line of sight. At urban intersections, vehicles may have only milliseconds of warning after a VRU emerges from occlusion, while rear cyclists or e-scooter riders may remain outside the primary sensor cone (Reyes-Muñoz & Guerrero-Ibáñez, 2022; Silva et al., 2025). Infrastructure-based sensing observes intersections continuously from elevated positions and can warn connected vehicles and personal devices before line-of-sight constraints arise (Mir et al., 2025; Zhang et al., 2025).

Two strands of work have advanced infrastructure-based VRU safety. The first develops sensors and detection algorithms: recent comprehensive surveys catalog dozens of infrastructure-side studies comparing cameras, thermal cameras, mmWave radar, mechanical and solid-state LiDAR, and other emerging modalities, fused at the data, feature, or decision level (Shang et al., 2025; Silva et al., 2025; Zhang et al., 2025). The second integrates infrastructure data into the vehicular network through Dedicated Short-Range Communications (DSRC), Cellular V2X (C-V2X), and 5G NR V2X, with documented vehicle-to-infrastructure signal-preemption trials (Schultz et al., 2022), connected-vehicle corridor pilots (Concas et al., 2021), a tutorial on the underlying standards (Castañeda Garcia et al., 2021), and a complementary survey of 5G location-based services for V2X (Mendes et al., 2025).

Despite the volume of this work, a clear gap remains. Existing reviews observe that no single sensor is sufficient to provide a robust perception system at urban intersections (Mir et al., 2025), and that current research disproportionately concentrates on autonomous vehicles equipped with extensive sensor technology rather than roadside-unit-based alternatives (Silva et al., 2025). Neither survey identifies a deployable architecture that couples multi-modal sensing with edge-side near-miss analytics, with bidirectional V2X and P2X messaging, and with adaptive signal control inside one intersection cabinet. Existing applied case studies either focus on a single sensor modality (Bhattarai et al., 2024; Lv et al., 2019) or demonstrate one of the framework layers in isolation (Fu et al., 2024; Pourhomayoun, 2024; Schultz et al., 2022). Multi-modal datasets that combine LiDAR, RGB, and thermal frames from real roadside deployments have only recently become public, with the R-LiViT dataset (Mirlach et al., 2025) the most prominent example.

This paper makes four contributions. First, it specifies an integrated four-layer framework for VRU protection at signalized intersections that combines (i) multi-modal roadside sensing with LiDAR, radar, RGB camera, and thermal camera, (ii) edge-based perception, prediction, and surrogate-safety analytics, (iii) bidirectional V2X and P2X messaging using Sensor Data Sharing Messages (SDSM) and Personal Safety Messages (PSM), and (iv) adaptive signal control. Second, it presents an empirical exposure-and-conflict-density case study built directly on the R-LiViT dataset using 53,319 detection annotations across 200 sequences, three intersections, and day-night recordings. Third, it provides a quantitatively grounded risk-decision table that links surrogate-safety thresholds drawn from the Swedish Traffic Conflicts Technique (Hydén, 1987) and modern conflict-to-crash transfer models (Tarko, 2018) to specific control actions. Fourth, it discusses honest limitations of treating image-plane proxies as approximations to true world-coordinate risk.

The remainder of the paper documents methodology and the R-LiViT case-study setup (Section 2), reviews relevant sensing, communication, edge, and surrogate-safety work (Section 3), specifies the framework (Section 4), presents the empirical case study (Section 5), and discusses implications, limitations, and future research (Section 6).

# 2. Materials and methods

## 2.1 Structured literature review

A structured narrative review was conducted across four bibliographic databases (Scopus, Web of Science, IEEE Xplore, and Google Scholar) using three concept groups combined with Boolean operators: (i) road-user terminology (vulnerable road user OR VRU OR pedestrian OR cyclist OR e-scooter), (ii) sensing and intelligent-transportation terminology (LiDAR OR radar OR sensor fusion OR V2X OR P2X OR edge computing OR smart intersection), and (iii) safety terminology (near-miss OR conflict OR surrogate safety OR TTC OR PET OR intersection safety). Searches prioritized peer-reviewed journal articles, conference proceedings, and authoritative technical reports, with selected high-relevance preprints retained when they directly supported the framework or case-study context. Foundational surrogate-safety works predating 2018, notably Hydén (1987), were retained for theoretical grounding. Studies focused exclusively on motorways, off-road environments, or in-vehicle driver-assistance without infrastructure relevance were excluded. After title and abstract screening, full-text review, and snowball expansion through key citations, 39 sources were retained and organized into seven thematic areas (Section 3, Table 2), with a small number of further sources cited for broader context: VRU exposure and crash patterns, single-sensor VRU detection, multi-sensor fusion, V2X and P2X communication, edge-based prediction and near-miss analytics, surrogate safety metrics, and adaptive signal control.

## 2.2 R-LiViT empirical case study setup

The R-LiViT dataset (Mirlach et al., 2025) was chosen because it is the first publicly released roadside LiDAR-Visual-Thermal dataset suitable for multi-modal exposure analysis at signalized intersections. R-LiViT comprises 200 sequences of 50 frames each captured at three German intersections, including 10,000 LiDAR point clouds and a subset of 2,400 temporally and spatially aligned RGB and thermal images jointly annotated for the 2D RGB-T benchmark. RGB images are 1280 by 720 pixels and thermal images are 640 by 512 pixels (FLIR ADK). Sequences are recorded in both daytime and nighttime conditions. The LiDAR benchmark provides annotations for seven object classes with per-object tracking identifiers; the 2D RGB-T benchmark provides annotations for eight object classes (person, bicycle, motorcycle, e-scooter, car, truck, bus, tramway) without per-object tracking identifiers.

For the descriptive case study, the publicly released 2D RGB-T annotation files were aggregated into a 53,319-row detection table indexed by sequence, frame, location, daytime flag, and class. Analyses in Python (pandas, numpy, matplotlib) covered per-class distribution and VRU share, per-frame VRU and vehicle density by day and night, spatial heatmaps of centroid coordinates, VRU-vehicle pairwise pixel-distance distributions at 50, 100, and 200 pixel thresholds, per-location density variability, and bounding-box size distributions as a proxy for sensor-relative distance. Because the 2D RGB-T benchmark lacks per-object track identifiers, pseudo-trajectories were not reconstructed, and pixel-space proximity is treated as an image-plane proxy for world-coordinate proximity.

## 3. Literature review

### 3.1 VRU exposure and crash patterns

Pedestrian injury severity in vehicle-non-motorist crashes is sharply non-linear in road environment, time of day, lighting, and pedestrian demographics, with tree-based machine-learning ensembles outperforming traditional logistic regression on 12,563 Florida vehicle-non-motorist crash records (Anowar, 2026). Posted speed limits affect crash risk primarily through their effect on operating speeds, with operating speed mediating approximately 48.7% of the speed-to-crash effect overall and a larger absolute magnitude of mediation on minor arterials, collectors, and local roads than on principal arterials (Wang et al., 2026). A recent Bayesian generalized extreme value model estimated 40.57 expected pedestrian crashes against 36 observed across nine Dhaka intersections, with non-linear link functions outperforming both stationary and linear-link alternatives (Anowar et al., 2025).

### 3.2 Single-sensor roadside detection

No single sensor modality adequately covers the multi-class, multi-lighting, multi-weather operating envelope at an urban intersection. Field comparisons at a four-way signalized intersection in Colorado Springs show radar achieving 400 m vehicle detection range, far exceeding its specified 275 m, but misclassifying VRUs because of small radar cross-section; LiDAR systems (Ouster OS1 around 100 m, Cepton X90 around 120 m) and cameras give superior VRU detection at shorter range (Mir et al., 2025). Thermal cameras, by leveraging heat signatures, give the strongest pedestrian detection at night and in low-visibility conditions (Choi et al., 2018). Multi-sensor surveys catalog the complementary strengths and limitations: cameras provide color and high resolution at low cost but degrade in glare and adverse weather; LiDAR generates accurate 3D point clouds but has eye-safety-limited range and rain or fog scattering; radar is robust to weather but produces noisy low-angular-resolution data; the mechanical-versus-solid-state distinction within LiDAR shifts the cost, resolution, and durability trade-off (Cui et al., 2022; Shang et al., 2025; Silva et al., 2025; Zhang et al., 2025).

Roadside LiDAR specifically has been demonstrated for proactive identification of vehicle-pedestrian and rear-end conflicts. A vehicle-pedestrian conflict identification system using roadside LiDAR with a rule-based speed-distance profile detected conflicts at signalized intersections without requiring crash records (Lv et al., 2019). Vehicle trajectory data from roadside LiDAR has been used to construct microscopic surrogate-safety indices that aggregate to rear-end risk indicators (Bhattarai et al., 2024). Automatic background filtering of roadside LiDAR point clouds can be performed at approximately 100 ms per frame after offline learning of the background mask, supporting real-time downstream detection (Wu et al., 2018), and elevated LiDAR mounted on traffic-light or street-lamp poles enables pedestrian monitoring with abnormal-activity detection (Guefrachi et al., 2024).

### 3.3 Multi-sensor fusion

Multi-sensor fusion strategies for VRU detection are typically organized along two orthogonal axes: data-level versus feature-level versus decision-level fusion, and early-fusion versus deep-fusion versus late-fusion stages in the network pipeline (Cui et al., 2022; Silva et al., 2025; Zhang et al., 2025). At the data level, early fusion preserves complete sensor information at the cost of high bandwidth and complex implementation. At the feature level, deep fusion compresses each modality into learned embeddings before combination, reducing bandwidth but losing some interpretability. Decision-level fusion combines per-sensor object lists, enabling flexible sensor addition and failure handling but losing intermediate information.

Recent intersection-side applications include drone-supervised radar-camera fusion at a Florida intersection, achieving a 22.6% average intersection-over-union (IoU) enhancement and a 30.4% overall ground-truth win rate across 19,612 ground-truth vehicle bounding boxes (Jahan et al., 2026), and a radar-camera fusion framework specifically designed around surrogate safety metrics including TTC and post-encroachment time (PET), which reports reductions of up to 67% in critical TTC events relative to camera-only baselines (Jahan et al., 2025). Late-fusion and early-fusion vehicle-infrastructure pipelines for camera-only and LiDAR-only modalities have been benchmarked on DAIR-V2X, the first large-scale vehicle-infrastructure cooperative 3D detection benchmark (Yu et al., 2022). Infrastructure-mounted machine vision has also been applied directly to VRU detection at intersections, reaching 82% mean average precision for pedestrians and cyclists in real time (Rahman et al., 2025), but only at the perception layer.

### 3.4 Communication: V2X, P2X, and 5G NR V2X

V2X communication includes vehicle-to-vehicle, vehicle-to-infrastructure, and vehicle-to-pedestrian variants and is increasingly extended to vehicle-to-network and vehicle-to-device exchanges in 5G New Radio profiles, collectively enabling cooperative awareness beyond any individual sensor's line of sight (Castañeda Garcia et al., 2021; Reyes-Muñoz & Guerrero-Ibáñez, 2022; Silva et al., 2025). Cellular V2X provides standards-aligned cellular implementations based on LTE and 5G New Radio, while DSRC offers low-latency peer-to-peer communication specifically designed for vehicular use (Castañeda Garcia et al., 2021; Mendes et al., 2025). Pilot deployments demonstrate practical viability: the Tampa THEA connected-vehicle pilot evaluated PSM alongside BSM, SRM, SSM, SPaT, MAP, and TIM message types across six use cases distributed over multiple downtown Tampa locations (Concas et al., 2021), and a Utah deployment showed that DSRC-equipped snowplows can request and obtain signal preemption from intersection-side roadside units (Schultz et al., 2022).

### 3.5 Edge-based prediction and surrogate safety

Edge computing places perception, prediction, and risk-assessment compute close to the sensor, reducing latency under stricter per-node resource constraints (Liu et al., 2019). Real-time fisheye-camera pipelines reach 40 fps with detection F1 of 0.988 and specificity above 0.99 after calibration and thin-plate-spline mapping (Huang et al., 2020). A pipeline-based prototype combines YOLO detection with Kalman-filter trajectory prediction on existing Los Angeles traffic-camera feeds (Pourhomayoun, 2024); a digital-twin prototype integrates edge perception with infrastructure-to-pedestrian MQTT messaging and was validated through CARLA hardware-in-the-loop testing (Fu et al., 2024); and in-vehicle dashcam edge prediction shows that the per-frame compute budget fits roadside cabinets of comparable scale (Dao & Zettsu, 2024). At the corridor scale, graph-learning models such as GCN-LSTM can predict traffic flow and speed and rank incident-response strategies far faster than microscopic simulation (Roy et al., 2026), complementing the intersection-level edge analytics emphasized here.

Surrogate safety measures infer crash risk from observed conflicts rather than from accumulated crash records. The Swedish Traffic Conflicts Technique introduced serious versus non-serious conflict zones tied to time-to-collision (TTC) and minimum distance (Hydén, 1987). Modern peer-reviewed treatment estimates the expected number of crashes from observed conflict counts using the Lomax distribution (Tarko, 2018), providing an empirical bridge from near-miss observation to expected-crash inference. Time-to-collision thresholds of approximately 1.5 s for vehicle-vehicle interaction and 2.5 s for vehicle-VRU interaction are widely used in the surrogate-safety literature and underpin the operational thresholds in Section 4 Table 1.

### 3.6 Adaptive signal control and AI-augmented analytics

Adaptive control in the context of emergency vehicle signal preemption uses DSRC on-board-unit broadcasts to trigger phase changes while a fixed minimum-green parameter protects pedestrians caught mid-crossing (Lee & Chiu, 2020); the same architecture extends to VRU detection inputs from infrastructure sensors. Recent surveys argue for AI-enhanced adaptive control that ingests roadside sensor streams continuously (Shang et al., 2025), and pedestrian-safety bibliometrics show a clear shift over the past five years toward collision-avoidance and intelligent-transportation clusters (Lee et al., 2025). A PRISMA-style review of VRU smart-technology studies confirms that camera, sensor, and tracker approaches dominate, with social-media, drone, and eye-tracking methods gaining ground (Parvez & Moridpour, 2025). Causal forests have begun to be applied to transportation safety operations (Tahmid et al., 2025), and the AV-pedestrian-interaction literature provides the behavioral-factor taxonomy for downstream policy modeling (Rasouli & Tsotsos, 2020).

## 4. Proposed framework

### 4.1 Architecture overview

Figure 1 presents the proposed four-layer roadside framework for VRU protection at signalized intersections. The layers are (i) Perception, comprising heterogeneous fixed sensors mounted on existing intersection infrastructure; (ii) Edge Computation, performing detection, tracking, prediction, and surrogate-safety analytics inside a cabinet-resident edge unit; (iii) Communication, exchanging Sensor Data Sharing Messages and Personal Safety Messages with connected vehicles and personal devices through C-V2X or DSRC; and (iv) Adaptive Control, modulating signal-phase timing and operating warning displays based on real-time risk estimates from the edge layer. The four layers operate as a continuous feedback loop, with control actions and warning events fed back into the perception layer to inform future risk estimates.

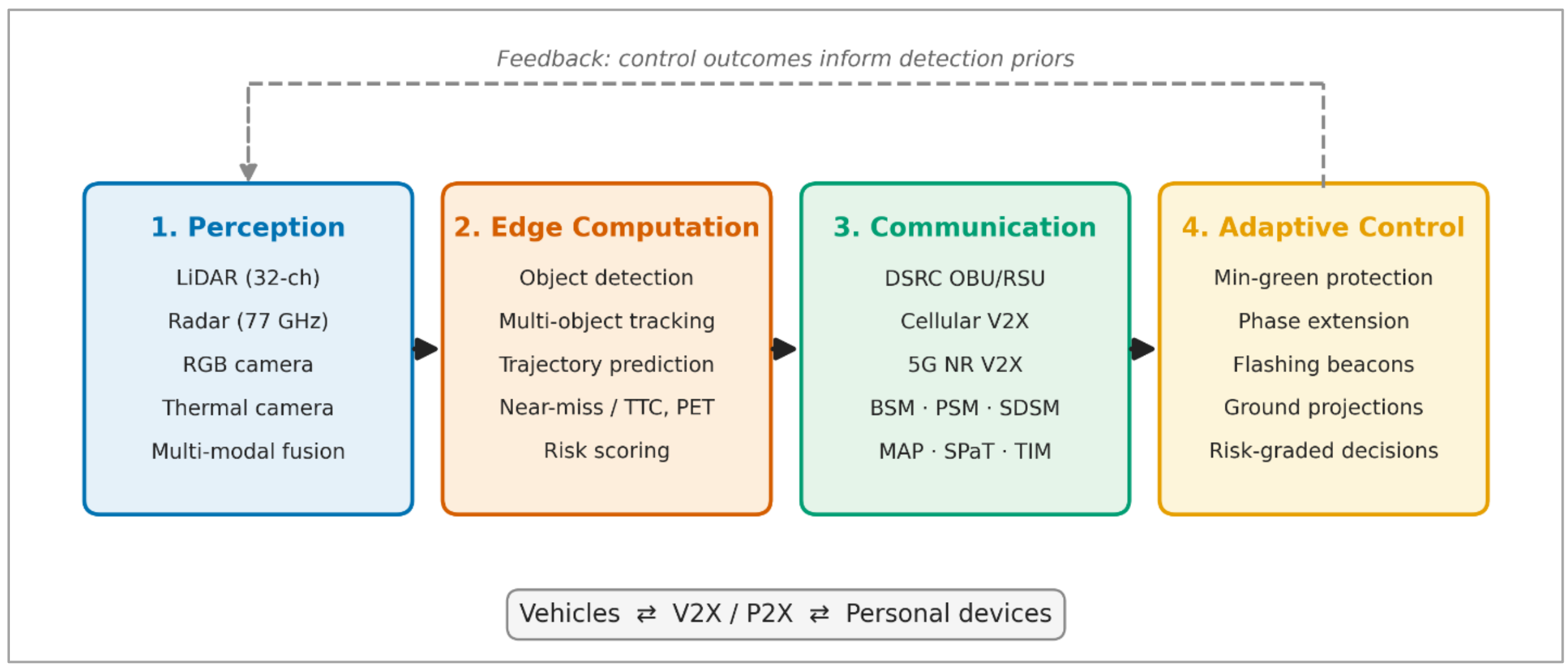

*Figure 1: Four-layer roadside framework for VRU protection.*

## 4.2 Perception layer

Four sensor modalities are deployed at each approach: a 32-channel mechanical LiDAR for 3D detection and trajectory geometry, a 77 GHz millimeter-wave radar for long-range velocity and weather-robust detection, an RGB camera for color and class disambiguation, and a thermal camera for night and low-visibility VRU detection. Each modality compensates for the others' documented weaknesses: cameras struggle with glare and night; thermal cameras are lower-resolution and lack texture; LiDAR is range-limited to approximately 100-120 m in the Ouster OS1 and Cepton X90 evaluated by Mir et al. (2025) and is degraded by heavy precipitation; radar misclassifies low-cross-section objects including pedestrians and cyclists (Cui et al., 2022; Mir et al., 2025; Silva et al., 2025; Zhang et al., 2025). Calibration follows the intrinsic-extrinsic-temporal workflow of Zhang et al. (2025), with targetless extrinsic methods preferred for ongoing maintenance.

Figure 2 shows the proposed deployment geometry at a representative four-leg signalized intersection (panel a) and the sensor placement detail (panel b). Sensors are mounted on the existing signal mast arms at heights of 4 to 6 m, with overlapping fields of view across the intersection box to ensure that each VRU is observable by at least two modalities at any moment.

## 4.3 Edge computation layer

The edge unit performs four computations inside a single intersection cabinet. First, per-sensor object detection runs YOLO-class convolutional neural networks on RGB and thermal streams, voxel-based 3D detectors on LiDAR point clouds, and clustering on radar point clouds. Second, multi-modal fusion combines per-sensor detections at the feature level for VRU classes (Cui et al., 2022) and at the decision level for vehicles. Third, short-horizon trajectory prediction uses a Kalman filter and a learned trajectory predictor on the fused tracks. Fourth, surrogate-safety analytics computes TTC and PET between every VRU-vehicle pair within the intersection box every frame.

Edge placement is preferred over cloud placement for latency (cabinet-to-cloud round trips exceed the action time available for VRU intervention), bandwidth (raw video at 30 fps across four cameras would saturate backhaul), and privacy (personally identifiable visual content remains local), addressing concerns flagged in recent reviews (Silva et al., 2025; Zhang et al., 2025).

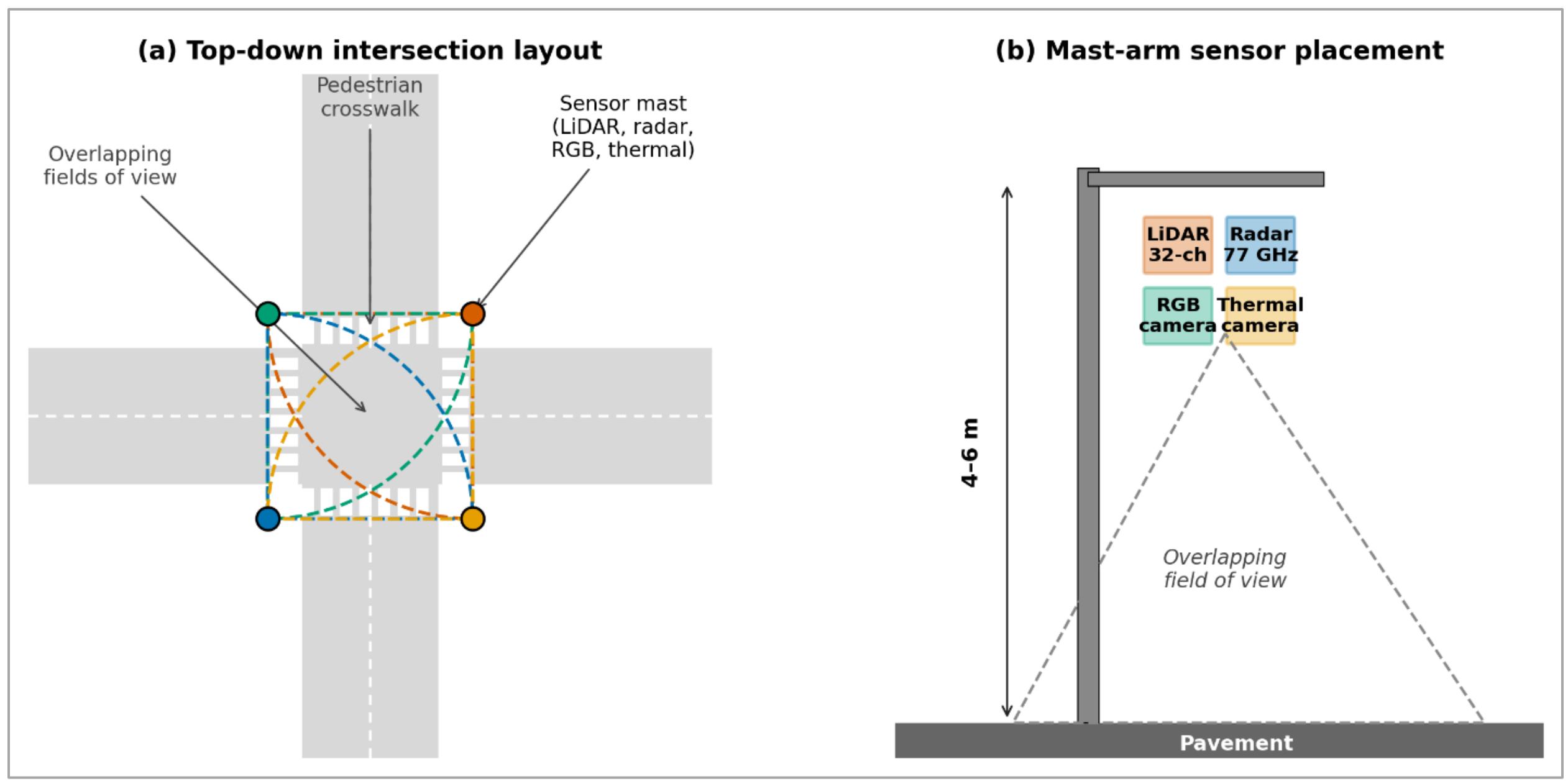

*Figure 2: Intersection deployment geometry.*

## 4.4 Communication layer

Two message types carry information out of the edge unit. Sensor Data Sharing Messages (SDSM) communicate detected objects (type, position, velocity, dimensions) from the infrastructure to connected vehicles, enabling the host vehicle to incorporate non-line-of-sight detections in its own planning. Personal Safety Messages (PSM) communicate VRU state to connected vehicles, either directly or via a relay through the infrastructure. Messages are exchanged over C-V2X PC5 sidelink (Castañeda Garcia et al., 2021) or DSRC, with cellular backhaul through 5G New Radio used for low-priority management traffic (Mendes et al., 2025). The Tampa THEA pilot has demonstrated PSM interoperability with the broader connected-vehicle message set including BSM, SRM, SSM, SPaT, MAP, and TIM across multiple downtown Tampa use cases (Concas et al., 2021), and Utah's snowplow signal-preemption deployment has demonstrated DSRC roadside-unit performance in active road operations (Schultz et al., 2022).

## 4.5 Adaptive control layer

Adaptive signal control extends pedestrian green when VRUs are in or approaching the conflict zone, modulates vehicle phase timing under high-risk conflict conditions, and operates flashing beacons and ground-projected warning displays for VRUs on approach. The minimum-green pedestrian-protection logic from emergency vehicle signal preemption (Lee & Chiu, 2020) provides a deployable baseline that the framework extends to VRU triggers using fused infrastructure detections, supported by recent AI-augmented adaptive signal proposals (Shang et al., 2025). Control logic follows a risk-graded decision table with TTC and PET thresholds from the surrogate-safety literature.

### 4.6 Risk-decision table

Table 1 maps measured TTC and PET values to specific framework responses. The thresholds are drawn from the foundational Swedish Traffic Conflicts Technique (Hydén, 1987), the Lomax-based crash-from-conflict estimator (Tarko, 2018), and modern roadside-LiDAR-based applications (Bhattarai et al., 2024; Lv et al., 2019).

*Table 1. Risk-decision table mapping TTC and PET thresholds to framework actions.*

| Risk band | TTC | PET | Edge response | Communication | Signal action |
|---|---|---|---|---|---|
| Low | > 2.5 s | > 1.5 s | Continuous tracking only | None | Nominal phase |
| Caution | 1.5 to 2.5 s | 0.8 to 1.5 s | Log conflict event | PSM advisory | Extend pedestrian green if active |
| Critical | 1.0 to 1.5 s | 0.5 to 0.8 s | Fast SDSM and PSM trigger | Emergency PSM, SDSM | Hold or extend conflicting phase; warning beacon |
| Imminent | ≤ 1.0 s | ≤ 0.5 s | Immediate alert | Broadcast emergency PSM | All-red, projector warning, fastest response |

### 4.7 Integration with smart-city analytics

The same edge stream that informs control decisions feeds longer-term safety analytics: aggregated near-miss counts at each location enable proactive hot-spot identification without waiting for crash records to accumulate (Bhattarai et al., 2024; Huang et al., 2020; Pourhomayoun, 2024). Aggregated conflicts can be evaluated using causal-inference approaches such as causal forests (Tahmid et al., 2025).

## 5. R-LiViT empirical case study

This section presents an empirical exposure-and-conflict-density case study using R-LiViT (Mirlach et al., 2025) annotation data, illustrating the rationale for the framework's multi-modal sensing layer and adaptive-deployment principle.

### 5.1 Dataset and aggregate distribution

R-LiViT is the first publicly released roadside dataset combining LiDAR, RGB, and thermal imagery with VRU-relevant object classes. The dataset spans three intersections in Germany and 200 sequences of 50 frames each, of which 2,400 RGB-T frames are jointly annotated across eight classes. The 53,319 annotations in the public 2D RGB-T files break down as 21,026 person, 4,142 bicycle, 456 motorcycle, and 389 e-scooter detections (26,013 VRUs in total) and 25,671 car, 810 truck, 511 bus, and 314 tramway detections. VRUs collectively account for approximately 49% of all road-user observations, putting VRU exposure on par with vehicle exposure rather than as a marginal share.

## 5.2 Multi-class VRU exposure

Figure 3, panel (a), shows the class distribution. The dominance of pedestrians within the VRU set (approximately 81%), combined with non-trivial shares of bicycles (approximately 16%), motorcycles (1.8%), and e-scooters (1.5%), justifies multi-class sensing at the perception layer. A single-class pedestrian detector would miss roughly one in five VRU observations.

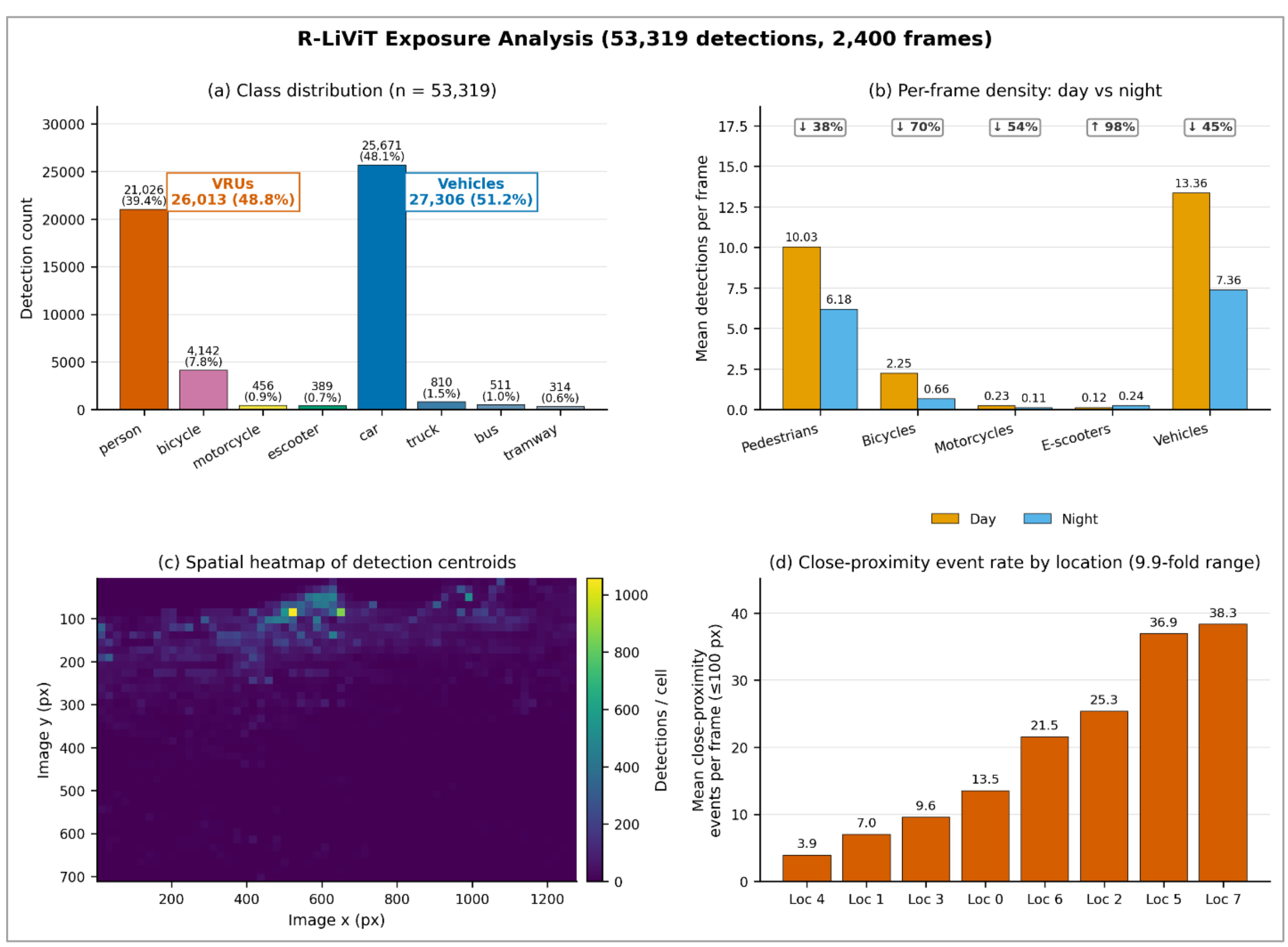

*Figure 3: R-LiViT exposure analysis.*

## 5.3 Day versus night density

Figure 3, panel (b), shows the per-frame density of pedestrians and vehicles separated by daytime and nighttime sequences. Daytime sequences contain on average 10.0 pedestrians per frame, while nighttime sequences contain 6.2 pedestrians per frame, a 38% reduction in absolute pedestrian density. Vehicle per-frame density drops from 13.4 to 7.4 vehicles per frame, about a 45% reduction. The relative VRU share of all road-user detections remains close to half across both lighting conditions, indicating that the safety problem is not primarily a daytime phenomenon and that thermal sensing for night detection is well motivated.

### 5.4 Spatial heatmap and co-presence

Figure 3, panel (c), shows the spatial heatmap of VRU and vehicle centroid coordinates in the image plane, aggregated across all sequences. VRUs cluster heavily at crosswalks and at sidewalk edges, while vehicles cluster within the road bed. Critically, 97.5% of all frames contain at least one VRU and at least one vehicle simultaneously, indicating that VRU-vehicle co-presence is the norm rather than the exception at the studied intersections.

### 5.5 Per-location variability

Figure 3, panel (d), shows per-frame close-proximity event counts across the eight unique locations at three intersections in the 2D RGB-T benchmark (three intersections, each split into day and night recordings; not every combination contributes equally). Close-proximity events (VRU-vehicle pairs within 100 pixels) per frame range from 3.9 at the quietest location to 38.3 at the busiest, an approximately 10-fold range. Per-frame VRU detection density varies more modestly over the same locations, from 5.6 to 15.9 per frame, a roughly 2.8-fold range. This implies that adaptive deployment, in which sensor configuration and edge analytics resources are sized per location, is important for cost-effective protection.

### 5.6 VRU-vehicle proximity distribution

Figure 4, panel (a), shows the distribution of pixel distances between every VRU-vehicle pair within a single frame, aggregated across 322,404 such pairs. The proximity distribution is heavy-tailed at short distances: 4.2% of pairs are within 50 pixels, 14.7% within 100 pixels, and 34.1% within 200 pixels. These pixel-space measures should be interpreted as image-plane proxies for true world-coordinate proximity; even so, the cumulative growth of close-pair counts justifies edge-side per-frame surrogate-safety computation rather than periodic batch analysis.

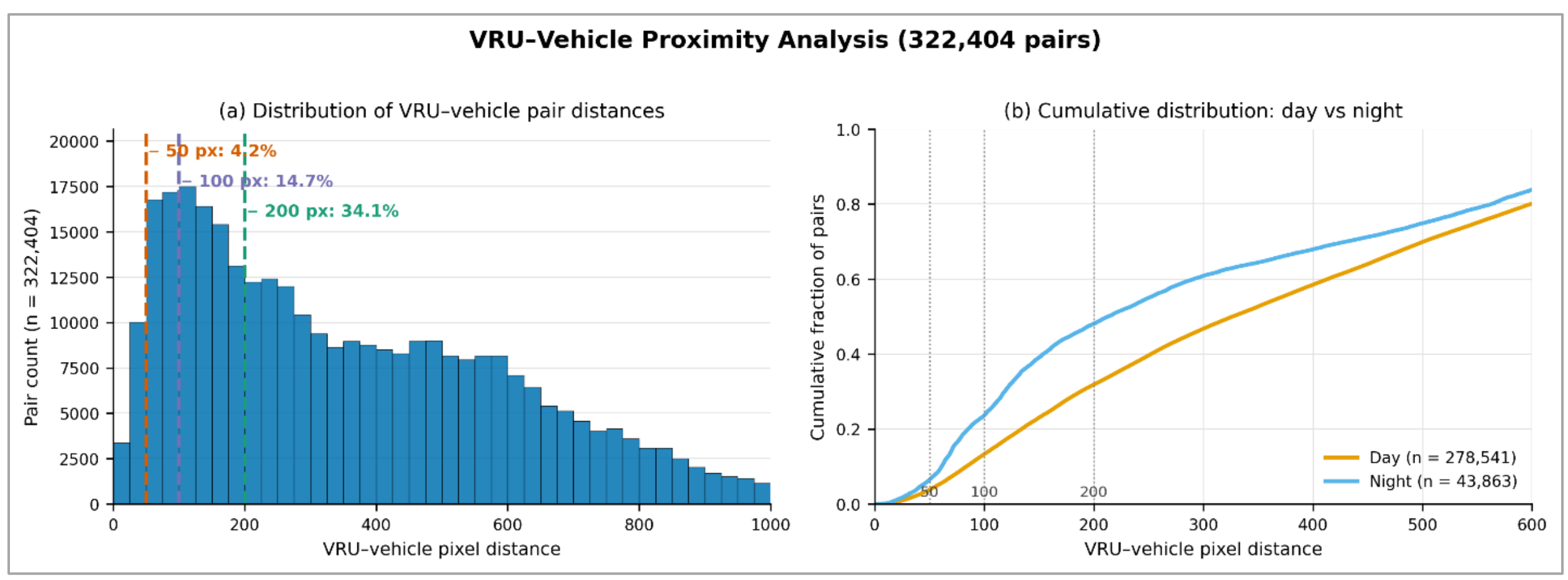

*Figure 4: VRU-vehicle proximity analysis.*

### 5.7 Day-versus-night proximity

Figure 4, panel (b), shows the cumulative distribution of VRU-vehicle pixel distances separately for day and night. While absolute close-pair counts drop at night because total VRU and vehicle densities fall, the cumulative share of close pairs at any given distance threshold is higher at night than by day, with 23.6% of night pairs within 100 pixels versus 13.3% by day. This pattern

suggests that close image-plane proximity remains important at night and supports the inclusion of thermal sensing and night-targeted analytics in the framework.

### 5.8 Bounding-box size analysis

A bounding-box size analysis (Appendix Figure A1) shows that 83.2% of pedestrian detections have an image-plane bounding-box area below 2,000 px², 70.4% of bicycle detections do, and 91.3% of e-scooter detections do. Because larger image-plane bounding boxes correspond to closer objects, this finding implies that VRUs are typically far from the camera and therefore close to the sensor-noise floor of a single visual modality, supporting the multi-modal sensing argument.

### 5.9 Mapping findings to framework decisions

The empirical findings map to specific framework decisions: (i) the 49% VRU share and multi-class composition justify multi-class detection; (ii) day-night differences justify thermal sensing; (iii) roughly 10-fold location variability justifies adaptive deployment; and (iv) high co-presence, heavy-tailed proximity, and small bounding boxes justify per-frame edge-side surrogate-safety computation with multi-modal redundancy.

Table 2 presents a cross-paper synthesis of the framework's seven thematic areas, mapping the retained literature to each area.

*Table 2. Cross-paper synthesis by thematic area.*

| Thematic area | Representative recent sources | Foundational source |
|---|---|---|
| VRU exposure and crash patterns | Anowar, 2026; Anowar et al., 2025; Pourhomayoun, 2024; World Health Organization, 2023 | Hydén, 1987 |
| Single-sensor VRU detection (LiDAR, radar, thermal, RGB) | Bhattarai et al., 2024; Choi et al., 2018; Guefrachi et al., 2024; Lv et al., 2019; Mir et al., 2025; Parvez & Moridpour, 2025; Rahman et al., 2025; Wu et al., 2018 | Cui et al., 2022 |
| Multi-sensor fusion | Jahan et al., 2025; Jahan et al., 2026; Yu et al., 2022; Zhang et al., 2025 | Silva et al., 2025 |
| V2X, P2X, and 5G NR V2X communication | Concas et al., 2021; Mendes et al., 2025; Schultz et al., 2022 | Castañeda Garcia et al., 2021; Reyes-Muñoz & Guerrero-Ibáñez, 2022 |
| Edge-based prediction and near-miss analytics | Dao & Zettsu, 2024; Fu et al., 2024; Huang et al., 2020; Pourhomayoun, 2024 | Liu et al., 2019 |
| Surrogate safety metrics | Bhattarai et al., 2024; Lv et al., 2019 | Hydén, 1987; Tarko, 2018 |
| Adaptive signal control, AI-augmented analytics, and AV interaction | Lee & Chiu, 2020; Lee et al., 2025; Shang et al., 2025; Tahmid et al., 2025; Wang et al., 2026 | Rasouli & Tsotsos, 2020 |

## 6. Discussion

### 6.1 Implications for intersection safety design

The R-LiViT exposure analysis provides quantitative support for design choices typically argued on conceptual grounds. Multi-modal sensing is justified because each sensor has documented weaknesses (Cui et al., 2022; Silva et al., 2025; Zhang et al., 2025) and because VRU exposure is bimodal in lighting, heavy-tailed in proximity, and varies roughly 3- to 10-fold by location. Edge-side analytics are justified by the per-frame co-presence rate, which makes batch-periodic analysis insufficient. The four-layer framework specified in Section 4 thus has empirical grounding beyond the typical conceptual reasoning. Intelligent transportation research also spans demand-side coordination, such as equilibrium pricing in multimodal mobility systems (Rafi & Guo, 2026); the present infrastructure is complementary, targeting supply-side road-user safety.

### 6.2 Limitations

Five limitations apply. First, R-LiViT records three German intersections; the modal composition (notable tramway presence) differs from U.S. or Australian deployments, and absolute density numbers should be tested before drawing region-specific conclusions. Second, the 2D RGB-T benchmark used here does not provide per-object track identifiers (the LiDAR benchmark does), so pseudo-trajectory continuity was not reconstructed, and pixel-distance proximity is an image-plane proxy rather than a world-coordinate measurement. Third, the proximity analysis is descriptive rather than predictive; converting close-proximity event counts into expected-crash estimates would require Lomax calibration to the specific intersection (Tarko, 2018). Fourth, the analysis is dataset-level only; the framework's adaptive control logic has not been tested in deployment, and the latency and reliability of the V2X stack are inferred from the literature rather than measured. Fifth, the framework does not address the substantial procurement, maintenance, and standardization economics of multi-sensor cabinets, which recent reviews identify as a non-trivial deployment barrier (Silva et al., 2025; Zhang et al., 2025). The framework's adaptive signal-control responses have not been evaluated in microsimulation; parameter-calibration methods for mixed traffic (Hoque et al., 2025) could support such evaluation before field deployment.

### 6.3 Future research

Five directions are worth pursuing. (i) End-to-end deployment in an instrumented intersection with full latency and reliability measurement of the SDSM and PSM messaging stack. (ii) Multi-region replication of the R-LiViT-style exposure analysis on U.S. and Australian datasets. (iii) Causal-inference evaluation of framework deployment using methods such as causal forests (Tahmid et al., 2025; Wang et al., 2026) to estimate per-layer conflict-reduction effects. (iv) Trajectory-level extension using the LiDAR portion of R-LiViT (which provides per-object track identifiers) or larger vehicle-infrastructure cooperative datasets (Yu et al., 2022). (v) Integration of large language model (LLM)-driven scene reasoning at the edge to handle the long tail of unusual VRU behavior (Silva et al., 2025).

## 7. Conclusion

This paper has presented an integrated four-layer framework for VRU protection at signalized intersections, combining multi-modal roadside sensing, edge-based analytics, V2X and P2X communication, and adaptive signal control. The framework is grounded in an empirical case study on the R-LiViT dataset, with 53,319 detection annotations across three German intersections showing that VRUs comprise approximately 49% of all road-user observations, that day-to-night pedestrian density drops 38% while VRU-vehicle pairs sit closer at night, that close-proximity event rates vary approximately 10-fold across the eight unique locations at three intersections, and that 83% of pedestrian bounding boxes are small in image space. These findings support multi-class multi-modal sensing, edge-side per-frame analytics, and adaptive context-sensitive deployment over uniform single-sensor installations. The risk-decision table maps surrogate-safety thresholds drawn from the Swedish Traffic Conflicts Technique and Lomax-based crash-from-conflict transfer to specific framework responses, providing a concrete starting point for cities upgrading signalized intersections to proactive infrastructure-mediated VRU protection.

## Acknowledgements

The authors thank the R-LiViT dataset team at XITASO GmbH, the Karlsruhe Institute of Technology, and LiangDao GmbH for releasing the dataset; R-LiViT was developed within the BMWK project VALISENS, funding code 19A22009E.

## Funding

This work received no specific grant from any funding agency in the public, commercial, or not-for-profit sectors.

## Conflicts of interest

The authors declare no conflicts of interest.

## Appendix

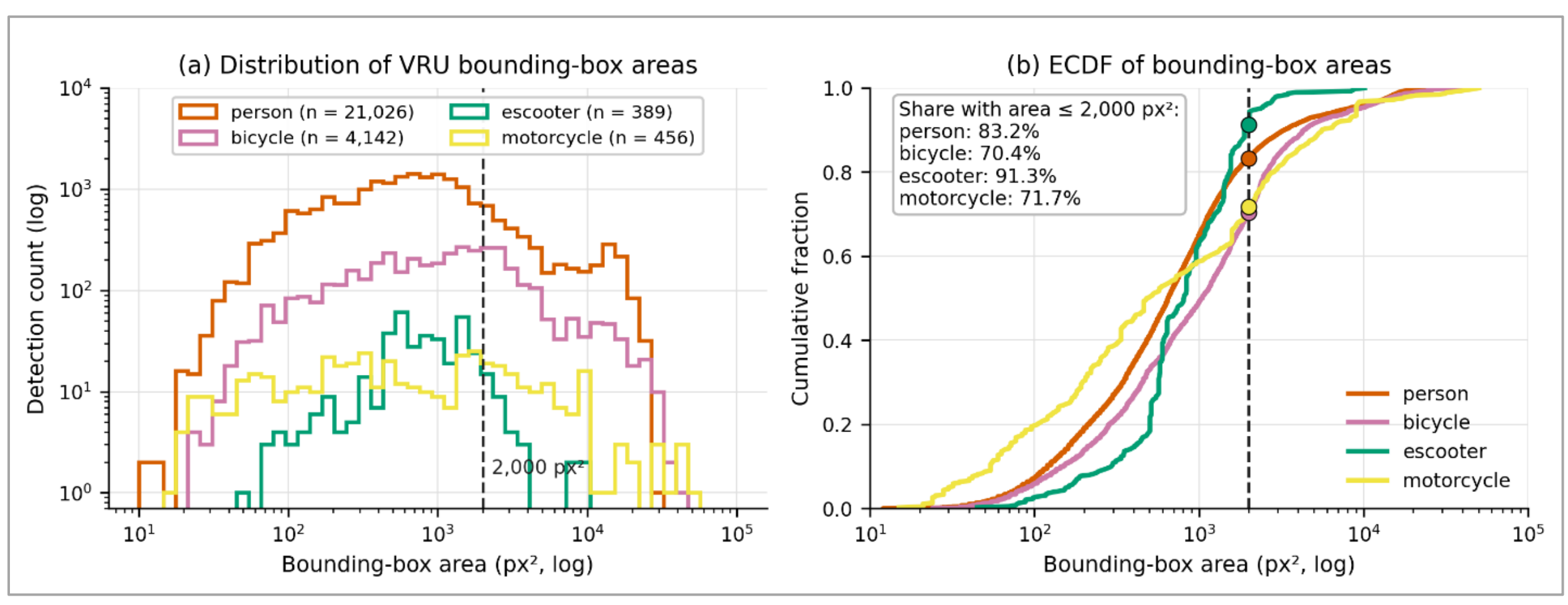

*Figure A1: Bounding-box size distribution by VRU class.*

## References

Anowar, P. (2026). Predicting injury severity in vehicle-non-motorist crashes: A comparative machine learning framework with interpretability analysis. engrXiv. https://doi.org/10.31224/6973

Anowar, P., Haque, N., Raihan, M. A., & Hadiuzzaman, M. (2025). Trajectory-based real-time pedestrian crash prediction at intersections: A novel non-linear link function for block maxima led Bayesian GEV framework addressing heterogeneous traffic condition. arXiv. https://arxiv.org/abs/2510.12963

Bhattarai, N., Zhang, Y., Liu, H., Pakzad, Y., & Xu, H. (2024). Proactive safety analysis using roadside LiDAR-based vehicle trajectory data: A study of rear-end crashes. *Transportation Research Record, 2678*(3), 772-785. https://doi.org/10.1177/03611981231182704

Castañeda Garcia, M. H., Molina-Galan, A., Boban, M., Gozalvez, J., Coll-Perales, B., Şahin, T., & Kousaridas, A. (2021). A tutorial on 5G NR V2X communications. *IEEE Communications Surveys & Tutorials, 23*(3), 1972-2026. https://doi.org/10.1109/COMST.2021.3057017

Choi, Y., Kim, N., Hwang, S., Park, K., Yoon, J. S., An, K., & Kweon, I. S. (2018). KAIST multi-spectral day/night data set for autonomous and assisted driving. *IEEE Transactions on Intelligent Transportation Systems, 19*(3), 934-948. https://doi.org/10.1109/TITS.2018.2791533

Concas, S., Kourtellis, A., Kamrani, M., & Dokur, O. (2021). *Connected vehicle pilot deployment program performance measurement and evaluation: Tampa (THEA) CV pilot Phase 3 evaluation report* (Report No. FHWA-JPO-20-829). ITS Joint Program Office, U.S. Department of Transportation.

Cui, Y., Chen, R., Chu, W., Chen, L., Tian, D., Li, Y., & Cao, D. (2022). Deep learning for image and point cloud fusion in autonomous driving: A review. *IEEE Transactions on Intelligent Transportation Systems, 23*(2), 722-739. https://doi.org/10.1109/TITS.2020.3023541

Dao, M.-S., & Zettsu, K. (2024). Near-miss accident prediction on the edge: A real-time system for safer driving. In *Proceedings of the 2024 International Conference on Multimedia Retrieval (ICMR '24)* (pp. 1165-1169). Association for Computing Machinery. https://doi.org/10.1145/3652583.3657623

Fu, Y., Turkcan, M. K., Anantha, V., Kostic, Z., Zussman, G., & Di, X. (2024). Digital twin for pedestrian safety warning at a single urban traffic intersection. In *2024 IEEE Intelligent Vehicles Symposium (IV)* (pp. 2640-2645). IEEE. https://doi.org/10.1109/IV55156.2024.10588544

Guefrachi, N., Shi, J., Ghazzai, H., & Alsharoa, A. (2024). Leveraging 3D LiDAR sensors to enable enhanced urban safety and public health: Pedestrian monitoring and abnormal activity detection. In *Proceedings of the 2024 46th Annual International Conference of the IEEE Engineering in Medicine and Biology Society (EMBC)*. IEEE. https://doi.org/10.1109/EMBC53108.2024.10782331

Hoque, I., Ananda, T. N., Anowar, P., Naz, A., & Murshed, M. N. (2025). Machine learning approach in calibrating VISSIM microsimulation model for mixed traffic conditions. *Journal of Engineering Science, 16*(1), 21-30. https://doi.org/10.3329/jes.v16i1.82663

Huang, X., Banerjee, T., Chen, K., Varanasi, N. V. S., Rangarajan, A., & Ranka, S. (2020). Machine learning based video processing for real-time near-miss detection. In *Proceedings of the 6th International Conference on Vehicle Technology and Intelligent Transport Systems (VEHITS 2020)* (pp. 169-179). SCITEPRESS. https://doi.org/10.5220/0009345401690179

Hydén, C. (1987). *The development of a method for traffic safety evaluation: The Swedish Traffic Conflicts Technique* (Bulletin No. 70). Lund Institute of Technology, Department of Traffic Planning and Engineering.

Jahan, I. A., Abdel-Aty, M., & Islam, Z. (2025). Impact of sensing modality on surrogate safety metrics: A radar-camera fusion framework for intersection monitoring. SSRN. https://ssrn.com/abstract=6486578

Jahan, I. A., Abdel-Aty, M., & Islam, Z. (2026). Drone-supervised multimodal sensor fusion for infrastructure-based vehicle detection in bird's eye view. *IEEE Internet of Things Journal, 13*(7), 13775-13786. https://doi.org/10.1109/JIOT.2026.3656347

Lee, M. B., Lee, C. T., Abas, M. A., & Chong, W. W. F. (2025). Advancing pedestrian safety in the era of autonomous vehicles: A bibliometric analysis and pathway to effective regulations. *Journal of Traffic and Transportation Engineering (English Edition), 12*(4), 772-794. https://doi.org/10.1016/j.jtte.2024.05.004

Lee, W.-H., & Chiu, C.-Y. (2020). Design and implementation of a smart traffic signal control system for smart city applications. *Sensors, 20*(2), 508. https://doi.org/10.3390/s20020508

Liu, S., Liu, L., Tang, J., Yu, B., Wang, Y., & Shi, W. (2019). Edge computing for autonomous driving: Opportunities and challenges. *Proceedings of the IEEE, 107*(8), 1697-1716. https://doi.org/10.1109/JPROC.2019.2915983

Lv, B., Sun, R., Zhang, H., Xu, H., & Yue, R. (2019). Automatic vehicle-pedestrian conflict identification with trajectories of road users extracted from roadside LiDAR sensors using a rule-based method. *IEEE Access, 7*, 161594-161606. https://doi.org/10.1109/ACCESS.2019.2951763

Mendes, B., Araújo, M., Goes, A., Corujo, D., & Oliveira, A. S. R. (2025). Exploring V2X in 5G networks: A comprehensive survey of location-based services in hybrid scenarios. *Vehicular Communications, 52*, 100878. https://doi.org/10.1016/j.vehcom.2025.100878

Mir, F., Sandhu, R., Young, S., Wang, Q., & Osborn, T. (2025). *The sensor dilemma in intelligent transportation systems: Evaluating radar, lidar and camera* (Report No. NREL/CP-5400-91797). National Renewable Energy Laboratory. https://www.nrel.gov/docs/fy25osti/91797.pdf

Mirlach, J., Wan, L., Wiedholz, A., Keen, H. E., & Eich, A. (2025). R-LiViT: A LiDAR-Visual-Thermal dataset enabling vulnerable road user focused roadside perception. In *Proceedings of the IEEE/CVF International Conference on Computer Vision (ICCV) 2025* (pp. 28375-28384). IEEE/Computer Vision Foundation. https://arxiv.org/abs/2503.17122

Parvez, M. S., & Moridpour, S. (2025). Application of smart technologies in safety of vulnerable road users: A review. *International Journal of Transportation Science and Technology, 18*, 285-304. https://doi.org/10.1016/j.ijtst.2024.07.006

Pourhomayoun, M. (2024). *Artificial intelligence for pedestrian and bicyclist safety: Using AI to detect near-miss collisions* (Report No. 24-34, Project 2350). Mineta Transportation Institute, San José State University. https://doi.org/10.31979/mti.2024.2350

Rafi, M. N. F., & Guo, Z. (2026). Multi-agent optimization of non-cooperative multimodal mobility systems. arXiv. https://doi.org/10.48550/arXiv.2601.03777

Rahman, M. A., Mammeri, A., & Metari, S. (2025). Intelligent infrastructure for enhancing vulnerable road user safety using machine vision technologies. *International Journal of Intelligent Transportation Systems Research, 23*, 1179-1196. https://doi.org/10.1007/s13177-025-00507-7

Rasouli, A., & Tsotsos, J. K. (2020). Autonomous vehicles that interact with pedestrians: A survey of theory and practice. *IEEE Transactions on Intelligent Transportation Systems, 21*(3), 900-918. https://doi.org/10.1109/TITS.2019.2901817

Reyes-Muñoz, A., & Guerrero-Ibáñez, J. (2022). Vulnerable road users and connected autonomous vehicles interaction: A survey. *Sensors, 22*(12), 4614. https://doi.org/10.3390/s22124614

Roy, S., Rashid, M. M., Rafi, M. N. F., Zhang, J., & Hasan, S. (2026). A novel graph learning-based approach to evaluate incident response strategies for integrated corridor management systems. *Transportation Research Part C: Emerging Technologies*, Article 105751. https://doi.org/10.1016/j.trc.2026.105751

Schultz, G. G., Lau, S. K., Shoaf, M., Bassett, D., & Eggett, D. L. (2022). *Analysis of using V2X DSRC-equipped snowplows to request signal preemption* (Report No. UT-22.14). Brigham Young University Department of Civil and Construction Engineering for the Utah Department of Transportation Research and Innovation Division.

Shang, B., Li, Y., Amin, A. G., Kamga, C., & Wei, J. (2025). AI-enhanced sensing for vulnerable road user safety at signalized intersections: A survey. *Procedia Computer Science, 265*, 350-357. https://doi.org/10.1016/j.procs.2025.07.191

Silva, R. M., Azevedo, G. F., Berto, M. V. V., Rocha, J. R., Fidelis, E. C., Nogueira, M. V., Lisboa, P. H., & Almeida, T. A. (2025). Vulnerable road user detection and safety enhancement: A comprehensive survey. *Expert Systems with Applications, 292*, 128529. https://doi.org/10.1016/j.eswa.2025.128529

Tahmid, M. M., Islam, Z., Abdel-Aty, M., Wang, C., & Ahsan, M. J. (2025). Estimating lane control signs (LCS) and variable speed limit (VSL) effects on expressway incident duration: A double machine learning causal forest approach. SSRN. https://ssrn.com/abstract=6199138

Tarko, A. P. (2018). Estimating the expected number of crashes with traffic conflicts and the Lomax distribution: A theoretical and numerical exploration. *Accident Analysis & Prevention, 113*, 63-73. https://doi.org/10.1016/j.aap.2018.01.008

Wang, C., Abdel-Aty, M., Zhai, S., Uddin, A. S. M. N., & Islam, Z. (2026). From prediction to explanation: A machine learning and causal mediation framework for roadway crash risk with connected vehicle data. *Transportation Research Part C: Emerging Technologies, 183*, 105479. https://doi.org/10.1016/j.trc.2025.105479

World Health Organization. (2023). *Global status report on road safety 2023*. https://www.who.int/teams/social-determinants-of-health/safety-and-mobility/global-status-report-on-road-safety-2023

Wu, J., Xu, H., Sun, Y., Zheng, J., & Yue, R. (2018). Automatic background filtering method for roadside LiDAR data. *Transportation Research Record, 2672*(45), 106-114. https://doi.org/10.1177/0361198118775841

Yu, H., Luo, Y., Shu, M., Huo, Y., Yang, Z., Shi, Y., Guo, Z., Li, H., Hu, X., Yuan, J., & Nie, Z. (2022). DAIR-V2X: A large-scale dataset for vehicle-infrastructure cooperative 3D object detection. In *Proceedings of the 2022 IEEE/CVF Conference on Computer Vision and Pattern Recognition (CVPR)* (pp. 21329-21338). IEEE. https://doi.org/10.1109/CVPR52688.2022.02067

Zhang, T., Cheng, L., Bang, T., Guo, L., Hajij, M., Cao, S., Harris, A., & Sartipi, M. (2025). Roadside sensor systems for vulnerable road user protection: A review of methods and applications. *IEEE Access, 13*, 62717-62738. https://doi.org/10.1109/ACCESS.2025.3558174